\documentclass[12pt]{article}

\usepackage{a4wide}
\usepackage{amssymb}
\usepackage{bm}
\usepackage{latexsym}
\usepackage{dcolumn}
\usepackage{amsfonts,amssymb}
\usepackage{graphicx,epsfig}
\usepackage{psfrag}
\usepackage{amsmath,amssymb}

\usepackage{bm}
\usepackage{latexsym}
\usepackage{mathrsfs}
\usepackage[T1]{fontenc}

\newcommand{\be}{\begin{equation}}
\newcommand{\ee}{\end{equation}}
\newcommand{\ba}{\begin{eqnarray}}
\newcommand{\ea}{\end{eqnarray}}
\newcommand{\ban}{\begin{eqnarray*}}
\newcommand{\ean}{\end{eqnarray*}}

\usepackage{hyperref}
\hypersetup{
    colorlinks=true,
    linkcolor=blue,
    filecolor=blue,      
    urlcolor=blue,
    citecolor=blue,
}
 
\begin{document}
\begin{titlepage}
\pagestyle{empty}
\baselineskip=21pt
\vspace{2cm}
\begin{center}
{\bf {\Large 
Strong gravity signatures in the polarization of gravitational waves\footnote{Essay received Honorable mention in Gravity Research Foundation essay competition-2019}
}}
\end{center}
\begin{center}
\vskip 0.2in
{\bf S. Shankaranarayanan}
\vskip 0.1in
{\it Department of Physics, Indian Institute of Technology Bombay, Mumbai 400076, India} \\
{\tt Email: shanki@phy.iitb.ac.in}\\
\end{center}

\vspace*{0.5cm}
\begin{abstract}
General Relativity is a hugely successful description of gravitation. However, both theory and observations suggest that General Relativity might have significant classical and quantum corrections in the \emph{Strong Gravity regime}. Testing the strong field limit of gravity is one of the main objectives of the future gravitational wave detectors. One way to detect strong gravity is through the \emph{polarization of gravitational waves}. For quasi-normal modes of black-holes in General Relativity, the two polarisation states of gravitational waves have the same amplitude and frequency spectrum. Using the principle of energy conservation, we show that, the polarisations differ for modified gravity theories. We obtain a diagnostic parameter for polarization mismatch that provides a unique way to distinguish General Relativity and modified gravity theories in gravitational wave detectors. 
\end{abstract}
\vfill\vfill
\end{titlepage}

\indent The principle of energy conservation is one of the cornerstones of modern physics. It has helped in making important discoveries, including, new particles. In the case of beta decay,  the electrons from beta decay were observed to have a continuum of energies ranging from zero to some maximum value. This observation signaled the possible violation of the conservation of energy in nuclear processes. Pauli suggested that if another unseen particle (neutrino) is emitted along with the electron that could take away part of the energy, then the energy is conserved in the process~\cite{Tomonaga}. In this essay, we apply this idea to distinguish General Relativity and modified theories of gravity in the strong gravity regime using gravitational wave detectors.  

The merger of two black-holes is a cataclysmic event, and it is still unclear whether General Relativity can be the exact description of gravity in the strong gravity regime. Both theory and observations suggest that General Relativity might have significant classical and quantum corrections in strong gravity regime~\cite{Modifiedtheories,MassiveGravity}. With the direct detection of gravitational waves from these cataclysmic events~\cite{GWObservations}, we can test gravity in the strong gravity regime using the advanced gravitational wave detection experiments~\cite{Futuredetectors}. There are many different ways to modify General Relativity in the strong gravity regime, and each model has different features. This leads to the question: Is there a unique signature that distinguishes General Relativity and modified gravity theories? To answer this question, we fall back on the basic principle of Physics --- \emph{the energy conservation}.

When two black holes merge to form another black-hole, the event horizon of the remnant black-hole is highly distorted and radiates gravitational waves until it settles down to an equilibrium configuration~\cite{GWavesreview}. In General Relativity, the distorted black-hole emits gravitatioanal waves with equal energy in the two polarization states~\cite{Chandrasekhar1984}. For modified theories of gravity --- due to additional propagating degrees of freedom --- the total emitted energy gets redistributed, and the two polarization states need not carry equal energy~\cite{SBSS}. The objective of this essay is to show that the principle of energy conservation can help us to distinguish between General Relativity and modified theories of gravity using gravitational wave detectors.  

Gravitational waves emitted by the distorted black-hole are quasi-normal modes, which are superposition of damped sinusoids~\cite{QNM}. Quasi-normal modes of a black-hole are independent of the disturbance that caused it and depends only on the mass and spin of the black-hole~\cite{QNM}. Thus, quasi-normal modes are the fingerprints of the final black-hole, and extracting their frequency and damping time allows for different tests of gravity~\cite{QNM,TestofGR}. More specifically, the complete understanding of the black-hole quasi-normal modes can place strong constraints on the mass of the graviton~\cite{MassiveGravity}, test Lorentz invariance violation in the gravitational sector, and modified gravity theories~\cite{Modifiedtheories}. 

Currently, these constraints are obtained by matching templates from General Relativity waveforms with observed data and introducing new parameters corresponding to modified gravity theories~\cite{GWObservations,GWavesreview}. However, the template matching technique will be inadequate for advanced detectors and the required accuracy to probe the modified General Relativity~\cite{Futuredetectors}. Thus, one has to identify other efficient ways to test deviations from General Relativity~\cite{Barack:2018yly}. 

In the rest of this essay, we show that the polarization provides a unique tool to distinguish between General Relativity and modified theories of gravity in the strong gravity regime. 
Consider a spherically symmetric space-time:
\be 
ds^2 = g(r) dt^2 - \frac{dr^2}{h(r)} - r^2 \left(d\theta^2 + \sin^2 \theta 
d\phi^2 \right) \, ,
\label{eq:SpSyST}
\ee
where $g(r)$ and $h(r)$ are arbitrary functions of $r$. At the linear order, the two modes of perturbations ($\Phi_1$ and $\Phi_2$) of the above space-time satisfy the following equations~\cite{Chandrasekhar1984}:
\begin{eqnarray}
\frac{d^2\Phi_i}{dr_*^2} + \left[ \omega^2-V_i(r_*)  \right] \Phi_i = 0 ~~~ & \mbox{where} & 
i= 1, 2 \, , \label{eq:GRper}
\end{eqnarray}
$r_* = \int dr/\sqrt{h(r) g(r)}$ is referred to as \emph{tortoise coordinate}, $\omega$ are the complex frequencies of the quasi-normal modes, and $V_i(r_*)$ are the potentials corresponding to the two modes ($\Phi_i$). The two potentials are related to each other via Darboux transformations. Hence,  the modes have the same reflection coefficients and carry equal energy~\cite{Chandrasekhar1984}. 

So, what happens in modified gravity theories? The field equations for a generic field-theoretic extension to General Relativity take the following compact form (taking $c = 1$)~\cite{Modifiedtheories}: 
\be
{\cal G}_{\mu\nu} = 8 \pi G \, T_{\mu\nu} \, ,
\ee
where ${\cal G}_{\mu\nu}$ is the modified Einstein tensor, $G$ is the 4-dimensional Newton's constant, and $T_{\mu\nu}$ is the stress-tensor of the minimally coupled matter fields to gravity. Irrespective of the gravitational field equations, we can demand the local conservation of the energy-momentum tensor $\nabla^{\mu} T_{\mu\nu}  = 0$. This condition is necessary to ensure that the equations of motion of the matter fields are unique. We then have the following generalized contracted Bianchi identity~\cite{LEnergyconservation}:
\be
\nabla^{\mu} {\cal G}_{\mu\nu} = 0 \, .  
\label{eq:Bianchiidentity}
\ee
To keep things transparent and understand the physical consequences of the generalized Bianchi identity, we consider two specific modified gravity theories --- $f(R)$ and Stelle gravity~\cite{Modifiedtheories,1978-Stelle-GRG}. We then generalize the results for a generic modified gravity theories. 

For $f(R)$ gravity where $f$ is an arbitrary, smooth function of the Ricci scalar ($R$), the modified Einstein tensor (${\cal G}_{\mu\nu}$) is given by~\cite{Modifiedtheories}:
\be 
\label{eq:ModEinsEq}
{\cal G}_{\mu \nu}  = f^{\prime}(R) \, R_{\mu \nu} - 
\nabla_{\mu} \nabla_{\nu} f^{\prime}(R) + g_{\mu \nu} \square f^{\prime}(R) 
-\frac{f(R)}{2} g_{\mu \nu}  \, ,
\ee
where $f'(R) = \partial f/\partial R$. The generalized Bianchi identity (\ref{eq:Bianchiidentity}) leads to:
\be 
\label{eq:BianchifR}
f^{\prime\prime}(R) \left( R_{\mu\nu} \nabla^{\mu} R \right) = 0 \, . 
\ee
For General Relativity, $f(R) = R$. Hence, $f^{\prime\prime}(R)$ vanishes and the above equation is trivially satisfied. However, $f^{\prime\prime}(R)$ is non-zero 
for modified gravity theories, hence, the generalized Bianchi identity (\ref{eq:BianchifR}) leads to four constraints on the Ricci tensor. While, General Relativity and $f(R)$ have four constraints on the field variables, the number of dynamical variables are different. For $f(R)$ gravity, unlike General Relativity, the trace of the field equation (\ref{eq:ModEinsEq}) is dynamical:
\be
\label{eq:Trace}
R \, f^{\prime}(R) +3 \, \square f^{\prime}(R) -2 \, f(R) =0
\ee
As a result, $f(R)$ gravity has 11 dynamical variables --- 10 metric variables ($g_{\mu\nu}$) and Ricci scalar ($R$). However, General Relativity has only 10 metric variables ($g_{\mu\nu}$). In other words, in $f(R)$, the scalar curvature $R$, plays a non-trivial role in the determination of the metric itself.  Since this extra degree of freedom is scalar, it can be treated as longitudinal mode~\cite{Futuredetectors}.  

Interestingly, the extra mode of propagation is a generic feature for any pure curvature modified theories of gravity containing only the higher-order Ricci scalar/tensor terms and without any additional matter fields. To see this, let us consider the following Stelle gravity action~\cite{1978-Stelle-GRG}:
\be 
S= \frac{1}{16 \pi G}  \int d^{4} x \sqrt{-g}\left(R - \alpha R^{2} + \beta R_{\mu \nu} R^{\mu \nu}  \right) \, , 
\ee
where $\alpha$ and $\beta$ are coupling constants. 
The modified Einstein tensor (${\cal G}_{\mu\nu}$) is
\begin{eqnarray}
{\cal G}_{\mu\nu} & = & G_{\mu\nu} + {\beta \left(-\frac{1}{2} R_{\rho \sigma} R^{\rho \sigma} g_{\mu \nu}-\nabla_{\nu} \nabla_{\mu} R-2 R_{\rho \nu \mu \sigma} R^{\sigma \rho}+\frac{1}{2} g_{\mu \nu} \square R+\square R_{\mu \nu}\right)} \nonumber \\ 
& + &  \, \alpha \left(\frac{1}{2} R^{2} g_{\mu \nu}-2 R R_{\mu \nu}-2 \nabla_{\nu} \nabla_{\mu} R+2 g_{\mu \nu} \square R\right) 
\end{eqnarray}
The above expression provides two key differences between any pure curvature modified gravity and General Relativity: First, like in $f(R)$ gravity, the trace of the modified Einstein equations (\ref{eq:ModEinsEq}) leads to dynamical equation of the Ricci scalar/tensor.  Second, the generalized Bianchi identity (\ref{eq:Bianchiidentity})  leads to four non-trivial constraints between Ricci tensor and Ricci scalar. Thus, any modifications to General Relativity will have at least \emph{one extra dynamical field} that plays non-trivial role in the determination of the metric itself.

The crucial point in the case of modified gravity theories is that as remnant black-hole settles down to an equilibrium state, some energy will be carried by the extra dynamical fields (say, $\Psi_{\rm new}$). This \emph{missing energy} --- like in the case of beta decay --- signals modifications to gravity. This leads us to the crucial question: How the missing energy can be used to distinguish General Relativity and modified gravity theories using gravitational wave detectors? 

As mentioned above, in the case of General Relativity, the two modes of perturbations satisfy
Eq. (\ref{eq:GRper}). However, for modified gravity theories, the two modes of perturbations for spherically symmetric space-times (\ref{eq:SpSyST}) --- that are related to the two polarizations detected by gravitational wave detectors --- satisfy the following relations~\cite{SBSS}:
\begin{eqnarray}
\frac{d^2\Phi_i}{dr_*^2} + \left(\omega^2-V_i\right)\Phi_i = 
S^{eff}_i\left(\Psi_{\rm new}\right) & &  i= 1, 2 
\label{eq:MGper}
\end{eqnarray}
where $S^{eff}_i(\Psi_{\rm new})$ are the effective source terms comprising of the new degrees of freedom. In general, $S^{eff}_1(\Psi_{\rm new}) \neq S^{eff}_2(\Psi_{\rm new})$.  For $f(R)$ theories it can be shown that $S^{eff}_2(\Psi_{\rm new})$ vanishes while $S^{eff}_2(\Psi_{\rm new})$ is non-zero~\cite{SBSS}. Thus, the two modes of perturbation in modified gravity theories \emph{do not} satisfy isospectral relation ($\Phi_1 \neq \Phi_2$) leading to an energetic inequality between the two observable modes in gravitational wave detectors. This energy inequality can be parameterized by a \emph{strong gravity diagnostic  parameter}~\cite{SBSS}:
\begin{eqnarray}
\Delta &=& \frac{\left|d_{t,\Omega}\Phi_1 \right|^2 - \left|d_{t,\Omega}\Phi_2\right|^2}{\left|d_{t,\Omega}\Phi_1 \right|^2 + \left|d_{t,\Omega}\Phi_2\right|^2} \label{SGdiagnostic} 
\end{eqnarray}
where the $ d_{t,\Omega} $ corresponds to derivative with respect to time and solid angle. In General Relativity, $\Phi_1 = \Phi_2$, hence, this parameter vanishes \emph{only} for General Relativity. For modified gravity theories, $\Phi_1 \neq \Phi_2$ and is non-zero.  In the case of $f(R)$ gravity, $\Delta \sim 10^{-7}$ for 
$10 M_{\odot}$ black-holes~\cite{SBSS}. In the next generation of gravitational-wave detectors (e.g. the Cosmic Explorer \cite{Evans:2016mbw}) the signal-to-noise ratio in the quasi-normal mode regime alone could be as large as ${\rm SNR} > 50$~\cite{QNMSNRbound}. With such detectors, the above diagnostic parameter can provide a unique signature for strong gravity. 

Although the above results are for spherically symmetric space-times, the diagnostic parameter is applicable also for rotating black-holes. For the rotating black-holes, the perturbation modes ($\Phi_1, \Phi_2$) need to be obtained numerically. This is currently being carried out for $f(R)$ theories of gravity.

In this essay, using the principle of energy conservation, we have developed a \emph{strong gravity diagnostic parameter} that can uniquely distinguish General Relativity and modified gravity theories in gravitational wave detectors. The gravitational wave detectors may be able to listen to the other octaves of gravitational wave symphony! 

\vspace*{10pt}

\noindent {\bf Acknowledgments} The author wishes to thank S. Bhattacharyya, Archana Pai, B. S. Sathyaprakash and Subodh R. Shenoy for discussions. The work is partially supported by Homi Bhabha Fellowship.



\end{document}